\documentclass[11pt]{llncs}
\usepackage{graphicx}
\usepackage{cite}
\usepackage{url}
\usepackage{times}
\usepackage{mathfont}
\def\log{\mathop{{\rm log}}}
\def\min{\mathop{\rm min}}
\def\max{\mathop{\rm max}}

\DeclareSymbolFont{AMSb}{U}{msb}{m}{n}
\DeclareSymbolFontAlphabet{\Bbb}{AMSb}
\def\Real{\ensuremath{\Bbb R}}

\mathcode`O="724F

\makeatletter
\def\hb@xt@{\hbox to }
\makeatother

\let\symdiff\triangle

\let\oldendproof\endproof
\def\endproof{\qed\oldendproof}

\begin{document}

\title{Algorithms for Media} 

\author{David Eppstein\inst{1} and Jean-Claude Falmagne\inst{2}}

\institute{University of California, Irvine,
Dept. of Information \& Computer Science\\
\and
University of California, Irvine,
Dept. of Cognitive Sciences\\
\email{\{eppstein,jcf\}@uci.edu}}

\maketitle   

\begin{abstract}
Falmagne recently introduced the concept of a {\em medium},
a combinatorial object encompassing  hyperplane arrangements, topological orderings, acyclic orientations, and many other familiar structures. We find efficient solutions for several algorithmic problems on media: finding short reset sequences, shortest paths, testing whether a medium has a closed orientation, and listing the states of a medium given a black-box description.
\end{abstract}

\section{Introduction}

Motivated by political choice theory, Falmagne~\cite{Fal-JMP-97} (see also Falmagne and Ovchinnikov~\cite{FalOvc-DAM-02})
recently introduced the concept of a medium,
a combinatorial object that also encompasses  hyperplane arrangements, topological orderings, acyclic orientations, and many other familiar combinatorial structures.

Formally, a {\em medium} consists of a set of {\em states} transformed by the actions
of a set of {\em tokens}, satisfying certain axioms.  That is, it is essentially a restricted type of deterministic finite automaton, without distinguished initial and final states.  Tokens may be concatenated to form {\em messages}
(words, in finite automaton terminology).   We use upper case letters to denote states, and lower case letters to denote tokens and messages; $Sw$ denotes the state formed by applying the tokens in message $w$ to state $S$.   A token $t$ is said to have a {\em reverse} $\tilde t$ if, for any two states $S\neq Q$, $St=Q$ iff $Q\tilde t=S$.  A message is said to be {\em inconsistent} if it contains some token and its reverse, and {\em consistent} otherwise.  A message $w$ is said to be {\em vacuous} if, for each token $t$, $w$ contains equal numbers of copies of $t$ and $\tilde t$.
A token $t$ is said to be {\em effective} for $S$ if $St\neq S$, and a message $w$ is be {\em stepwise effective} for $S$ if each successive token in the sequence of transformations of $Sw$ is effective. A medium is then defined to be a system of states and tokens satisfying the following axioms:

\begin{enumerate}
\item Each token has a unique reverse.
\item For any two distinct states $S$, $Q$, there exists a consistent message $w$ with $Sw=Q$.
\item If message $w$  is stepwise effective for $S$, then  $Sw=S$ if and only if $w$ is vacuous.
\item If $Sw=Qz$, $w$ is stepwise effective for $S$, $z$ is stepwise effective for $Q$, and
both $w$ and $z$ are consistent, then $wz$ is consistent.
\end{enumerate}

The following are examples of media:

\begin{description}
\item[{\bf Permutations.}]
The set of permutations of $n$ items form the states of a medium, with a token $t_{xy}$ for each ordered pair $xy$ of items that replaces an adjacent pair $yx$ in a permutation by the pair $xy$, or leaves the permutation unchanged if no such pair exists.  The reverse of $t_{xy}$ is $t_{yx}$.

\medskip
\item[{\bf Topological orderings.}]
For any directed acyclic graph $G$, one can define a medium with states that are the topological orderings of $G$, and the same swap operations $t_{xy}$ as above for each pair of unrelated vertices in the graph.  When $G$ has no edges, we get the permutation medium on the vertices of $G$.

\medskip
\item[{\bf Acyclic orientations.}]
Let $G$ be an undirected graph, define a state to be an acyclic orientation of the edges of $G$,
and for any ordered pair $xy$ of adjacent vertices define a token $t_{xy}$ that reorients edge $(x,y)$ from $x$ to $y$, if the resulting orientation is acyclic, and leaves the orientation unchanged otherwise.  The result is a medium.  When $G$ is complete, it reduces to the permutation medium on the vertices of $G$.

\medskip
\item[{\bf Hyperplane arrangements.}]
Let $C$ be a convex region in $\Real^d$, and
$A$ be a hyperplane arrangement in $C$. Then the convex cells of $A$ form the states
of a medium, with one token $t_h$ for each halfspace $h$ bounded by a hyperplane in
$A$.  If a cell $S$ is included in $C\setminus h$ and shares a facet with a cell $S'$ included
in $h$, then $St_h=S'$; otherwise $St_h=S$.  In the special case where $C=\Real^n$, with Cartesian coordinates $x_i$, and $A$ is the arrangement of hyperplanes $x_i=x_j$ for the edges
$(i,j)$ of an $n$-vertex graph, this is isomorphic to the acyclic
orientation medium described above.  We can also realize the medium of topological
orderings by using hyperplanes $x_i=x_j$ for all $i$ and $j$, with $C$ consisting
of the points in $\Real^d$ satisfying $x_i\le x_j$ for each arc $(i,j)$ in the given
DAG.

\medskip
\item[{\bf Well-graded families of sets.}]
A family
$\mathcal W$ of subsets of a set $X = \cup \mathcal W$ is {\em well-graded} ~\cite{DoiFal-DM-97} if, for any two distinct sets $S$ and
$Q$ in $\mathcal W$, there exists a sequence $S= S_0,S_1,\ldots,S_k=Q$ in $\mathcal W$ such
that
$|S_{j-1}\symdiff S_{j}| = 1$ for $1\leq j\leq k=|S\symdiff Q|$. (Here, $\symdiff$ denotes the set-theoretic symmetric difference.) Falmagne and
Ovchinnikov~\cite{FalOvc-DAM-02} show that such a family can be cast as a medium, the states of
which are the sets in $\mathcal W$. To each $x$ in $X$ are associated two tokens $i_x$ and
$d_x$, defined respectively by: $Si_x = S\cup \{x\}$ if $S\cup \{x\} \in \mathcal W$, and
$Si_x = S$ otherwise; and $Sd_x = S\setminus \{x\}$ if $S\setminus \{x\} \in \mathcal W$, and
$Sd_x = S$ otherwise. It is easily verified that the set of states $\mathcal W$ and the
collection of tokens $i_x$ and $d_x$ satisfy the axioms of a medium. Notice that for each
$x$ in $X$, the tokens $i_x$ and $d_x$ are mutual reverses. As suggested by the remaining
examples  in this section, this structure subsumes many special cases.

\medskip
\item[{\bf Downward closed set families.}] Let $\mathcal F$ denote a
family of subsets of a set $X=\cup\mathcal F$ having the property that $\mathcal F$ contains any subset of any of its members.
For instance, the independent sets in a
graph or matroid define such a family.
Then $\mathcal F$ is well-graded and so can be represented as a medium.

\medskip
\item[{\bf Binary trees.}]
One can encode a binary tree as a set of integers, that
give the heap numbers of nodes in the tree: the root has number $0$, and the left and
right children of a node with number $i$ have numbers $2i+1$ and $2i+2$.  Let
$\mathcal F$ be the family of sets encoding binary trees with at most $k$ leaves;
then $\mathcal F$ is well-graded and can be represented as a medium.
We can similarly represent the trees with height at most $k$.

\medskip
\item[{\bf Antimatroids and convex geometries.}]
%%If a family $\mathcal F$ of
%%subsets of a set $X$ is well-graded, then the family $\overline {\mathcal F}$ of complements of
%%sets in $\mathcal F$ with respect to $X$ is also well-graded.
As noted by Doignon and
Falmagne~\cite{DoiFal-99},
%%this implies that
well graded families closed under union are dual to {\em antimatroids}  or {\em convex geometries} in the sense of Jamison and Edelman~\cite{EdeJam-GD-85} (see also
\cite{KorLovSch-91}). Such structures can thus also be represented as media.
\end{description}

Notice that the media constructed in the last four examples
are endowed with a natural `orientation' in that the tokens are either adding or removing some
element. Orientations of media were examined in \cite{FalOvc-DAM-02} and we use them also in this paper; in some sense oriented media are equivalent to well-graded set families (Lemma~\ref{well-graded-from-medium}). The permutation
and  arrangement examples are given by Falmagne and Ovchinnikov~\cite{FalOvc-DAM-02} who
also  provide several additional examples of media, including families of partial orders,
interval orders, semiorders, and biorders.  Previous work~\cite{Fal-JMP-97,FalOvc-DAM-02} has
studied the combinatorics and applications of media.  In this paper, we provide a first study
of algorithms for media.

\section{New Results}

In the study of deterministic finite automata, an important concept that arises is that of a {\em reset sequence}~\cite{Moo-AS-56,Gin-JACM-58}: an input word $w$ such that applying this word to a state $S$ yields a transformed result $Sw$ that is independent of $S$.  Thus, if we imagine an ensemble of copies of the automaton, each initially running in a distinct initial state, the application of this word will cause the ensemble to be synchronized to a common state.  Reset sequences have been applied, for example, in the design of devices for orienting machine parts~\cite{Nat-FOCS-86}.  If a reset sequence exists for an $n$-state automaton, there is one with $O(n^3)$ symbols, which can be constructed in polynomial time; {\v{C}}ern{\'y}~\cite{Cer-MFCSAV-64} conjectured a tight bound of $(n-1)^2$ on the length of the shortest reset sequence for any automaton, but this is only known for special classes of automata~\cite{Dub-RAIRO-98,Epp-SJC-90,Rys-TCS-97}.
Our first result is a tight bound of $n-1$ on the length of the shortest reset sequence for a medium. We also describe an algorithm for finding a sequence with this length in time linear in the length of a simple description of the medium.

Second, we study medium-theoretic concepts of distance.  We define length functions on tokens and use them to define a natural concept of shortest paths in a medium.  We show how to find single source shortest paths in linear time, and all pairs shortest paths in time quadratic in the number of states.  We define a notion of the complement of a state in a medium, generalizing set-theoretic complementation in well-graded set families, and show how to find complements more quickly than all pairs shortest paths.

Third, we consider the concept of a {\em closed orientation}, introduced and studied by Falmagne and Ovchinnikov~\cite{FalOvc-DAM-02}.  We describe a near-linear time for testing whether an orientation is closed, and a polynomial time algorithm for finding a closed orientation if one exists.

Finally, we study {\em black box} oracle-based definitions of media, similar to the black box groups introduced by Babai and Szemer\'edi~\cite{BabSze-FOCS-84} and since studied extensively in computational group theory.  We use a modified version of the {\em reverse search} procedure of Avis and Fukuda~\cite{AviFuk-DAM-96} to show that the states of a black box medium can be listed in time and space per state that is polynomial in the number of tokens of the medium, generalizing previously known algorithms for listing the states of the example media described in the introduction.

\section{Preliminaries}
\label{preliminaries}

In order to perform algorithms analysis with media, we need some conventions for input format and size parameters.  For any medium, we let $n$ denote the number of states, and $\tau$ denote the number of tokens.  One simple format for describing a medium in a computer would be as an $n\times\tau$ matrix the cells of which contain the results of applying each token to each state.
However, for many media, an even more concise representation is possible: define an {\em adjacency list} representation of a medium to be an $n$-dimensional array, where, for any state $S$,
array cell $A[S]$ contains a list of the pairs $(t,Q)$ for which $St=Q\neq S$.  That is, we list only the effective transitions of the medium.  We let $m$ denote the number of pairs
listed in all cells of this array.
As an example for which $m$ is much smaller than $n\tau$, consider a one-dimensional hyperplane arrangement formed by $n-1$ points on a line: $n\tau=2n(n-1)$ while $m=2(n-1)$.

We now recall some definitions from Falmagne and Ovchinnikov~\cite{Fal-JMP-97,FalOvc-DAM-02}.  A consistent message $w$ which is stepwise effective for state $S$ is called a {\em straight path} from $S$ to $Sw$; we say more briefly that $Sw=Q$ is a straight path.   Axiom 2 in the definition of a medium can be strengthened to the statement that for any $S$ and $Q$ there is a straight path $Sw=Q$.   For any state $Q$ in a medium, the {\em content} of $Q$ ($\hat Q$, for short) is defined to be the set of tokens that can occur in messages $w$ that determine straight paths $Sw=Q$.

\begin{lemma}[Falmagne~\cite{Fal-JMP-97}]
\label{content-identifies-state}
If $Sw=Q$ is a straight path, then $\hat Q\setminus\hat S$ consists of exactly the tokens in $w$.
\end{lemma}

\begin{lemma}[Falmagne~\cite{Fal-JMP-97}]
\label{content-is-partition}
For any token $t$ and any state $Q$, exactly one of the two tokens $t$, $\tilde t$ belongs to $\hat Q$.
\end{lemma}

Denote the set of tokens of a medium by $\mathcal T$.
Falmagne and Ovchinnikov~\cite{FalOvc-DAM-02} define an {\em orientation} of a medium to be a partition $\{\mathcal T^+, \mathcal T^-\}$ of $\mathcal T$ into positive and negative subsets such that, for each token $t$, exactly one of $t$ and $\tilde t$ is positive.  By Lemma~\ref{content-is-partition}, the partition
$\{\hat Q,\mathcal T\setminus\hat Q\}$ forms an orientation, which we call the {\em content orientation} of $Q$.
For any orientation $\{\mathcal T^+, \mathcal T^-\}$
we call set $\hat Q^+=\hat Q\cap\mathcal T^+$ the {\em positive content} of $Q$.
As we mentioned in the introduction, a well-graded family of sets $\mathcal F$
provides a set-theoretic definition of a medium
$(\mathcal S_{\mathcal F},\mathcal T_{\mathcal F})$ equipped with a natural orientation $\{T_{\mathcal F}^+=\{i_x\mid x\in \cup\mathcal F\},T_{\mathcal F}^-=\{d_x\mid x\in \cup\mathcal F\}\}$.  Indeed, we have the following equivalence between well-graded set families and oriented media, for which we omit the proof:

\begin{lemma}
\label{well-graded-from-medium}
For any orientation  $\{\mathcal T^+, \mathcal T^-\}$ of a medium $(\mathcal S,\mathcal T)$,
the family $\mathcal F$ of positive contents of states in $\mathcal S$ is well-graded,
and $(\mathcal S,\mathcal T)$ is isomorphic to $(\mathcal S_{\mathcal F},\mathcal T_{\mathcal F})$;
that is, there exist two bijections $f:\mathcal S\mapsto\mathcal S_{\mathcal F}$ and
$g:\mathcal T\mapsto\mathcal T_{\mathcal F}$ such that
$$St=Q\quad\Longleftrightarrow\quad
f(S)g(t)=f(Q),
\quad\quad(S,Q\in\mathcal F,t\in\mathcal T).$$
Moreover the orientation $\{\mathcal T^+, \mathcal T^-\}$ matches the natural orientation
$\{T_{\mathcal F}^+,T_{\mathcal F}^-\}$ of the medium for $\mathcal F$, in that $t\in\mathcal T^+$ if and only if $g(t)\in\mathcal T_{\mathcal F}^+$.
\end{lemma}

\begin{corollary}
Any medium with $\tau$ tokens has at most $2^{\tau/2}$ states.
\end{corollary}

As a consequence of this, a factor of $\log n$ in the running time of an algorithm on a medium is always preferable to a factor of $\tau$.  Even more preferable would be a factor of $m/n$, due to the following lemma:

\begin{lemma}
\label{medium-density}
In any medium, $m\le n\log_2 n$.
\end{lemma}

\begin{proof}
Since media can be defined by well-graded set families,
we prove more generally that, in any family $\mathcal{F}$ of $n$ sets, the number of unordered pairs of sets
$(P,Q)$ with $|P\symdiff Q|=1$ is at most $\frac12n\log_2 n$; each such pair
contributes two effective transitions to~$m$.
This bound appears to be well known (compare Lemma~3 of Matou\v{s}ek~\cite{Mat-uso})
but for completeness we prove it here.

As a base case for our induction on $n$, a family of one set has no such pairs.
Choose $x\in\cup\mathcal{F}$, belonging to some but not all sets in $\mathcal{F}$, and divide $\mathcal{F}$ into two subfamilies
$\mathcal{F}_x$ and $\mathcal{F}_{\bar x}$ where $\mathcal{F}_x$ consists of the members of $\mathcal{F}$ that contain $x$ and $\mathcal{F}_{\bar x}$ consists of the remaining members of $\mathcal{F}$;
let $|\mathcal{F}_x|=a$ and $|\mathcal{F}_{\bar x}|=b$.
Then each set $P$ in $\mathcal{F}$ is involved in at most one pair $(P,P\symdiff\{x\})$ with one set in each subfamily, so the number of such pairs is at most $\min(a,b)$.  If a pair $(P,Q)$ is not of this form, then
both $P$ and $Q$ belong to the same subfamily.  So, if $M(n)$ denotes the maximum number of pairs defined by any $n$-set family, we have a recurrence
$$M(n)\le\max_{a+b=n} M(a)+M(b)+\min(a,b)$$
which solves to the desired bound.
\end{proof}

%% Media are also related to the {\em accessible set families} defined by
%% Korte, Lov\'asz, and Schrader~\cite{KorLovSch-91}: a family is accessible
%% if from each set in the family we can form another set in the family by
%% removing one element.  A well-graded set family containing the empty set
%% is accessible; in particular, this is true for the family of negative contents
%% of states for a content orientation of any medium.  There are accessible
%% set families that are not well-graded and so do not form media.  However,
%% as indicated earlier, an {\em antimatroid} (accessible set family closed
%% under union) is well-graded: in any antimatroid $\mathcal F$, if
%% $B\setminus A\neq\emptyset$, there exists an $x\in B$ such that
%% $A\cup\{x\}\in\mathcal F$~\cite{KorLovSch-91}, so we can find a well-graded
%% sequence of element insertions connecting $A$ to $A\cup B$ and
%% symmetrically a sequence of deletions connecting $A\cup B$ to $B$.
%% The complements of the sets in an antimatroid form a {\em convex
%% geometry}~\cite{KorLovSch-91}, and since complementation preserves
%% well-gradedness, convex geometries are also well graded.  The family
%% of sets used in the introduction to define binary trees of height at most $k$
%% forms an antimatroid, although the family of sets defining trees with
%% at most $k$ leaves does not.

\section{Reset Sequences}

Recall that a reset sequence for a medium is a message $w$ such that $Sw=Qw$ for every two states $S$ and $Q$.  We now describe an algorithm for efficiently finding short reset sequences.

Given an oriented medium, one can construct a directed graph $G$ the vertices of which correspond to the states of the medium and the arcs of which represent transitions: draw an arc from $S$ to $Q$ whenever $St=Q\neq S$ for some positive token $t$.  Axiom 3 of the definition of a medium is easily seen to imply that $G$ is acyclic.

\begin{lemma}
\label{content-unique-sink}
The graph $G$ constructed from the content orientation of state $Q$ has $Q$ as its unique sink.
\end{lemma}

\begin{proof}
If $Qt=S$, $S{\tilde t}=Q$, so ${\tilde t}$ is a member of $\hat Q$ and the edge in $G$ connecting $S$ with $Q$ is oriented as an arc from $S$ to $Q$; thus $Q$ is a sink.
Any other state $S$ has a straight path $Sw=Q$, each step of which corresponds to an arc oriented from $S$ towards $Q$, so $S$ can not be a sink.
\end{proof}

\begin{theorem}
If we are given as input the adjacency list representation of a medium, then in time $O(m)$ we can find a reset sequence $w$ for the medium with length $|w|\le n-1$, such that for any state $S$ of the medium, $Sw=Q$.
\end{theorem}

\begin{proof}
Construct the graph $G$ as in Lemma~\ref{content-unique-sink}; it has $n$ vertices and $m/2$ arcs. In time linear in the size of $G$, find a topological ordering $S_0, S_1, \ldots, S_{n-1}=Q$ of the graph.  Then output $w=p_{S_0}p_{S_1}\ldots p_{S_{n-2}}$.

To prove that $w$ is a reset sequence, we show by induction that, if $w_i$ denotes the $i$-symbol prefix of $w$, and $S$ is any state of the medium, then $Sw_i=S_j$ for some $j\ge i$.
As a base case, if $i=0$, the statement is vacuous.
Otherwise, by induction, $Sw_{i-1}=S_{j'}$ for some $j'\ge i-1$.
Note that $w_i=w_{i-1}t$ for the positive token $t=p_{S_{i-1}}$,
so $Sw_i=S_{j'}t$.  Since $t$ is positive, $S_{j'}t=S_j$ for $j\ge j'$, with equality only if $t$ is ineffective for $j'$.  But $t$ was chosen to be effective for $S_{i-1}$, so $j\ge i$.
\end{proof}

We remark that better bounds on the reset sequence length, in terms of $\tau$ instead of $n$, may be possible: for instance, one can form a reset sequence with length $(\tau/2)^2$ simply by concatenating $\tau/2$ copies of a word $w$ containing all positive tokens for the content orientation of a state $Q$.  However for some media $(\tau/2)^2$ may be larger than $n-1$. Since $\tau/2\le n-1$, it is natural to hope that a reset sequence of length $\tau/2$ always exists, but this is not true; for instance there is no such reset sequence for a medium with six states and six tokens, formed by the well-graded family of all 1- and 2-element subsets of a 3-element set.

\section{Distances and Complements}

It is natural to define distances in a medium by assigning lengths to the medium's tokens. We define a {\em length function} for a medium $(\mathcal S, \mathcal T)$ to be a function $\lambda:\mathcal T\mapsto\Real$, satisfying the constraint that $\lambda(t)+\lambda(\tilde t)\ge 0$ for every token~$t$.  We define the {\em length} of a path $Sw=Q$ to be the sum of the lengths of the effective tokens on the path, and the {\em distance} from state $S$ to state $Q$ to be the length of the shortest path $Sw=Q$.  Note that these distances satisfy the triangle inequality but need not be symmetric or nonnegative.  Then it is not hard to see that every straight path is a shortest
path, so shortest path computation amounts to finding a straight path between the given states.  Distances to a single state~$Q$ from every other state may be found in time $O(m)$ by following paths in the DAG of Lemma~\ref{content-unique-sink}, and single source shortest paths can be found by reversing the edges of this DAG.  Somewhat less trivially, we can also speed up the time for all pairs shortest paths relative to the time for similar computations in graphs:

\begin{theorem}
If we are given as input the adjacency list representation of a medium, then in time $O(n^2)$ we can build an $n\times n$ table that lists, for each two states $S$ and $Q$, the distance from $S$ to $Q$, as well as a token $t$ such that $St$ is effective and $t$ belongs to a straight path from $S$ to $Q$.
\end{theorem}

\begin{proof}
We expand the representation of the medium to an explicit $n\times\tau$ table of transitions that allows us to test in constant time whether a token is effective for a given state.
We then perform a depth first traversal of the states of the medium, maintaining as we do a data structure that allows us to compute the table entries for each traversed state $Q$.  The depth first traversal itself makes at most $2n-3$ transitions before all states are reached, and this sequence of transitions can be constructed in time $O(m)$.
The data structure that we maintain in the traversal consists of the following components:
\begin{itemize}
\item A doubly-linked list $L$ of pairs $(t,\Lambda_t)$, where each $t$ is a token in the content of $Q$ and $\Lambda_t$ is a pointer to a linked list described below.
\item A pointer from each state $S\neq Q$ to the first pair $(t,\Lambda_t)$ in $L$ for which $St$ is effective.
\item Linked lists $\Lambda_t$ for each pair $(t,\Lambda_t)$, listing the states pointing to that pair. 
\end{itemize}

The pointers from each state $S$ to the associated pair $(t,\Lambda_t)$ provide half of the information we are trying to compute: an effective transition on a straight path to $Q$ from each other state $S$. We record this information for $Q$ when the traversal first reaches $Q$.
The other information, the numeric distances to $Q$, can easily be computed in time $O(n)$ for each $Q$ by traversing the tree formed by the effective transitions $St$.

We initialize the data structure by creating an empty list $\Lambda_t$ for each token in the content of the initial state $Q$, listing the pairs $(t,\Lambda_t)$ in an arbitrary order, and sequentially searching this list to initialize the pointer for each state $S\neq Q$.
It remains to describe how to update the data structure as we perform each transition
$Qt=Q'$ of the depth first traversal.  Token $\tilde t$ belongs to the content of $Q$, so prior to the transition it is listed as part of some pair $(\tilde t,\Lambda_{\tilde t})$ in $L$.  We remove this pair from $L$, and append a new pair $(t,\Lambda_t)$ to the end of $L$, where $\Lambda_t$ is a new empty list.  We must then recompute the pointers from each state $S$ that had previously been pointing to $(\tilde t,\Lambda_{\tilde t})$; we do so by sequentially searching list $L$, starting at the position of the deleted pair $(\tilde t,\Lambda_{\tilde t})$, for the first pair $(t',\Lambda_{t'})$ such that $St'$ is effective, and appending $S$ to $\Lambda_{t'}$.
We set the pointer for $Q$ to the pair $(t,\Lambda_t)$ without searching, since $t$ will be the only
token in $\hat Q'$ that is effective for $Q$.

We finish the proof by analyzing the time used by this algorithm.
List $L$ initially contains $\tau/2$ pairs, and each traversal step appends a new pair to $L$; therefore the total number of pairs added to $L$ over the course of the algorithm is at most
$\tau/2+2n-3=O(n)$. The most expensive part of the algorithm is the sequential searching to find a new effective pair.  For each state $S$, the sequence of sequential search steps never revisits a position in $L$, so the total number of steps of sequential searching over the course of the algorithm is at most $n(\tau/2+2n-3)=O(n^2)$.  Computing numeric distances also takes a total of $O(n^2)$ time, and the other data structure update steps take only constant time per traversal step.  Therefore, the total time for the algorithm is $O(n^2)$.
\end{proof}

From the table constructed above, one can construct a straight path $Sw=Q$ for any two states $S$ and $Q$, in time $O(|w|)$, by repeatedly using the table to find effective tokens in the path.

Related to distance is the concept of complementation in a medium.
We define the {\em complement} of a state $S$ to be a state $Q$ such that $S$ and $Q$ have disjoint contents.  If a complement exists, it is unique, and is the farthest state from $S$ for any length function.  We can test whether a medium has a complement for all of its states in $O(m)$ time, very simply, by searching for the complement $Q$ of a single state $S$, and then performing two parallel traversals through the medium starting from these two states, maintaining as an invariant that the state visited by the second traversal is complementary to the state visited by the first traversal. This idea can be extended to a method for finding all complementary pairs, somewhat less efficiently but still faster than the all pairs shortest paths algorithm above.

\begin{theorem}
In time $O(n\tau)$ we can find all complementary pairs of states in a medium.
\end{theorem}

\begin{proof}
As in the previous algorithm, we expand the medium representation to one of size $O(n\tau)$ that allows for fast tests of the effectiveness of a transition as well as for determining whether or not a token belongs to the content of a given state.
We then perform a depth first traversal of all states $Q$.  At each step of the traversal we maintain the following data:
\begin{itemize}
\item A state $S$
\item A list $L$ of the tokens in $\hat Q\cap\hat S$
\end{itemize}
If $L$ is empty, $S$ must be the complement of $Q$.  Otherwise, if $Q$ has a complement $S'\neq S$, then $L$ must contain each token on a straight path $S'w=S$, and is nonempty.
To perform a step $Qt=Q'$ of the traversal, we remove $\tilde t$ from $L$ (if it was there),
and add $t$ to $L$ (if it belongs to the content of $S$).
Then, as long as we can find a token $t'\in L$ with $S\tilde t'$ effective,
we replace $S$ by $S\tilde t'$ and remove $t'$ from $L$.
We search for $t'$ by scanning $L$ after each change of $S$.
If we reach a state $S$ with $L$ empty, $S$ is the complement of $Q$;
otherwise, $Q$ has no complement.
We scan $L$ once per change of $Q$, and once per change of $S$.
Since $Q$ follows a depth first traversal, it changes $O(n)$ times.
Each change of $S$ is accompanied by the removal of an item from $L$,
and items are added to $L$ only when $Q$ changes, so the number of changes to $S$ is also bounded by $O(n)$.  Thus, the total time is $O(n\tau)$.
\end{proof}

\section{Closed Orientations}

Falmagne and Ovchinnikov~\cite{FalOvc-DAM-02} define a {\em closed orientation} of a medium to be one in which, whenever positive tokens $t$ and $t'$ are both effective for any state $S$,
the messages $tt'$ and $t't$ are stepwise effective for $S$.
For instance, the natural orientation of an antimatroid is closed.
Closed orientations have several useful properties: for instance, any state can be transformed to any other state by a stepwise-effective {\em canonical message} consisting of a sequence of positive tokens followed by a sequence of negative tokens; additionally, like the content orientations used in the previous section, the DAG formed from a closed orientation has a unique sink.  A naive algorithm for testing whether an orientation is closed (testing each triple $S$, $t$, $t'$) would take time $O(n\tau^2)$.  We improve this with the following simple observation, which can be viewed as a strengthening of Lemma~\ref{medium-density} showing that the number of effective tokens is small on a per-state basis instead of only when totalled.

\begin{lemma}
Let $\{\mathcal T^+,\mathcal T^-\}$ be a closed orientation of a medium.
Then for any state $S$ there are at most $\log_2 n$ positive tokens effective for $S$.
\end{lemma}

\begin{proof}
Let $E$ denote the set of positive tokens effective for $S$.
Then it follows from the definition of a closed orientation
that, for each $E'\subset E$, there is a distinct state $S_{E'}=Sw_{E'}$ where $w_{E'}$ is formed by concatenating the tokens in $E'$.  Thus the medium contains at least
$2^{|E|}$ states, from which the result follows.
\end{proof}

Note however that the number of negative effective tokens may be as large as $n-1$.

\begin{theorem}
\label{test-closed}
We can test whether an orientation of a medium is closed in time $O(m\log n)$.
\end{theorem}

\begin{proof}
In time $O(m)$ we can determine the set of effective positive tokens for each state.
If any state has more than $\log_2 n$ tokens, we can immediately determine that
the orientation is not closed.  Otherwise, we test each triple $S$, $t$, $t'$ where $t$ and $t'$ are selected only from the positive tokens effective for $S$.
The number of pairs $S$, $t$ involved in such triples is at most $m$,
and each such pair forms at most $\log_2 n$ triples,
so the number of tests is $O(m\log n)$.
\end{proof}

The method above is non-constructive, in the sense that it may determine that an oriented medium is not closed without finding an explicit triple $S$, $t$, $t'$ violating the definition of a closed medium.  With some additional care we can make it constructive:

\begin{lemma}
In any closed oriented medium, let $p(S)$ denote the number of positive effective tokens for state $S$, and let $t$ be a positive token.  Then $p(St)\ge p(S)-1$.
\end{lemma}

\begin{proof}
By the definition of a closed medium, every positive effective token for $S$ other than $t$ itself must also be effective for $St$.
\end{proof}

\begin{theorem}
If an oriented medium is not closed, we can find a triple $S$, $t$, $t'$ violating the definition of a closed medium in time $O(m\log n)$.
\end{theorem}

\begin{proof}
We can compute $p(S)$ for each state $S$ in total time $O(m)$.
If some state $S$ and positive token $t$ have $p(St)<p(S)-1$,
then we can compare the lists of effective tokens for $S$ and $St$ in time
$O(\tau)$ and find a token $t'$ that is effective for $S$ but ineffective for $St$;
$S$, $t$, and $t'$ form the desired triple.

Next, if $p(St)\ge p(S)-1$ is always true, but some state has more than $\log_2 n$ positive effective tokens, we can follow a sequence of positive effective transitions from that state until we find a state $S$ with exactly $1+\lfloor\log_2 n\rfloor$ positive effective tokens.
Let $w=t_1\ldots t_k$ be a message formed by concatenating all such tokens; we can perform a breadth first search to find a minimal subsequence $w'$ of $w$ such that $Sw'$ is not stepwise effective.  Such a subsequence must exist because (by the axioms of a medium) each stepwise effective subsequence of $w$ transforms $S$ to a distinct state, and there can be at most $n$ states found altogether, but there are $2^k>n$ distinct subsequences of $w$.
This search takes time $O(2^k)=O(n)$.
Let $w'=vtt'$ for some message $v$ and tokens $t$ and $t'$;
then $Sv$, $t$, $t'$ form the desired triple.
Finally, if all states have at most $\log_2 n$ positive effective tokens,
we can perform the same search for violating triples used in Theorem~\ref{test-closed}.
\end{proof}

Next, we consider the problem of finding closed orientations for unoriented media.
The simplest medium without a closed orientation is one with six states and six tokens, formed by the well-graded family of all 1- and 2-element subsets of a 3-element set.

\begin{theorem}
If we are given an unoriented medium, we can find a closed orientation, if one exists, or determine that no such orientation exists, in time $O(m\tau)$.
\end{theorem}

\begin{proof}
We reduce the problem to an instance of 2-SAT~\cite{AspPlaTar-IPL-79}, by creating a Boolean variable for each token in the medium, where the truth of variable $t$ in a truth assignment for our instance will correspond to the positivity of token $t$ in an orientation of the medium.  By adding clauses $t\vee \tilde t$ and $\neg t\vee\neg\tilde t$, we guarantee that exactly one of the two variables $t$ and $\tilde t$ is true, so that the truth assignments of the instance correspond exactly to orientations of the medium.  We then add further clauses $\neg t\vee\neg t'$ for each triple $S$, $t$, $t'$ such that $St$ and $St'$ are effective but $Stt'$ is not stepwise effective. In an orientation corresponding to a truth assignment satisfying these clauses, $t$ and $t'$ can not both be positive, so triple $S$, $t$, $t'$ will not violate the definition of a closed orientation.  The 2-SAT instance can be constructed in time $O(m\tau)$, and solved in time $O(\tau^2)$. 
\end{proof}

An alternative method for finding closed orientations is based on the observation that any orientation with a unique sink $Q$ must be the content orientation of $Q$.  We can test each content orientation using Theorem~\ref{test-closed}.  However this does not seem to lead to a method more efficient than the one described above.

\section{Black Box Media}

A medium may have exponentially many states, relative to its number of tokens, making storage of all states and transitions infeasible.  However, for many media, a single state can easily be stored, and state transitions can be constructed algorithmically.  Thus, we would like to design algorithms for media that do not depend on storing an explicit list of states and state transitions.  However, we would like to do this in as general way as possible, not depending on details of the state representation.  This motivates an oracle-based model of media, similar to the black box groups
from computational group theory~\cite{BabSze-FOCS-84}.

We define a {\em black box medium} to be a representation of a medium consisting of the following parts:
\begin{itemize}
\item A list of the tokens of the medium.
\item A procedure {\tt transitionFunction} that takes as input the representation of a state $S$ and a token $t$, and produces as output the representation of the state $St$.
\item The representation of a single state $S_0$.
\end{itemize}
We require each state to have a unique representation, so that states can be compared for equality, however the details of the state representation and transition function are unavailable.

For example, all of the well-graded set family examples discussed in the introduction could be implemented by using a bitmap to represent a set, with the transition function calling another function to determine whether a given set belongs to the family, but this type of {\em independence oracle} allows more computations than we consider here~\cite{KarUpfWig-STOC-85,KarUpfWig-FOCS-85}.  The hyperplane arrangement example could be implemented by using linear programming to determine whether a token produces an effective transition.

As before, we use $\tau$ to denote the number of tokens of a black box medium, and
$n$ to denote the number of states.  Since any medium has a finite number of states, we can bound the amount of memory required to store a single state by a parameter~$s$.  We denote by $T$ the amount of time required per call to {\tt transitionFunction}.  In this context, we consider ``polynomial time'' to mean time that is polynomial in $\tau$ and $T$, and
``polynomial space'' to mean a space bound that is polynomial in $\tau$ and $s$.

One of the most basic computational problems for a black box medium is to find all of its possible states.  As with any deterministic finite automaton, one could list all states by performing a depth first traversal of the state transition graph; however, if the medium is large enough to make an adjacency list representation infeasibly large, then it would also likely be infeasible to store the set of already-visited nodes needed to allow depth first search to avoid repetition of states.  As we now show, it is possible instead to traverse all states of a black box medium, using an amount of storage limited to only a constant number of states.

The main idea behind our traversal algorithm is to use a modified version of the {\em reverse search} procedure of Avis and Fukuda~\cite{AviFuk-DAM-96}.  The reverse search procedure allows one to list all the states of a general state transition system, as long as one can define {\em canonical paths}: sequences of transitions from each state to an initially given state $S_0$ such that the union of the transitions on all the paths forms a spanning tree of the state space, the {\em canonical path tree}. If one can quickly determine whether a transition belongs to the canonical path tree, then one can list the states by traversing this tree, ignoring the transitions that do not form edges in the tree.
In our case, canonical paths can be constructed from the content orientation of the initial state $S_0$.  If we have an ordered list of the positive tokens of the content orientation, then a step in a canonical path from state $S$ can be found by taking the first token in the list that is effective on $S$.  The difficulty, however, is that we are not given the content orientation, and can not construct it without searching the medium, the task we are trying to solve.  Fortunately, we can interleave the construction of the orientation and the enumeration of states, in a way that allows the reverse search algorithm to proceed, taking polynomial time per state and using polynomial space.

\begin{figure}[p]
\begin{verbatim}
def mediumStates(s0,tokens,transitionFunction):
    '''List all states of a black box medium.
    The input is a single starting state s0, a list of tokens, and a procedure that
    performs the transition function (state,token)->state of a medium.
    The output is a simple generator yielding all states of the medium.'''

    positiveTokens = {}
        # dictionary of tokens that can occur on a straight path to s0.
        # since we need to store something in positiveTokens[t],
        # we use it to store the inverse token of t, but this inverse
        # is not actually used by the search algorithm.
    
    def step(state):
        '''Return a token that would take the given state on a straight path to s0.'''
        for token in positiveTokens:
            neighbor = transitionFunction(state,token)
            if neighbor != state: return neighbor
        raise ValueError("bad medium: unable to find a positive step from " + repr(state))
    
    state = s0
    yield state
    tokenSequence = iter(tokens)
    while 1:
        try:
            token = tokenSequence.next()
            if token in positiveTokens: continue
            neighbor = transitionFunction(state,token)
            if neighbor == state: continue
            
            # here with an effective nonpositive token.
            # make sure its inverse is listed as positive so we can call step()
            for invtok in tokens:
                if transitionFunction(neighbor,invtok) == state:
                    positiveTokens[invtok] = token
                    
            # check whether state->neighbor is the reverse of a step
            if step(neighbor) == state:
                state = neighbor
                yield state
                tokenSequence = iter(tokens)
            
        except StopIteration:
            # here after exhausting all tokens for the current state.
            if state == s0:
                return

            # we need to backtrack in the search.
            # the parent state can be found with step() but we also need
            # to find where the parent's search left off in the token sequence.
            parent = step(state)
            tokenSequence = iter(tokens)
            for token in tokenSequence:
                if transitionFunction(parent,token) == state:
                    break # found correct point in token sequence
            state = parent
\end{verbatim}
\caption{Python implementation of reverse search for black box media states.}
\label{Python-reverse-search}
\end{figure}

\begin{theorem}
\label{list-medium-states}
If we are given a black box medium, we can list all states of the medium
in time $O(n\tau^2T)$ and space $O(s+\tau)$.
\end{theorem}

\begin{proof}
We perform the reverse search procedure described above, building a list of positive tokens in the content orientation of $S_0$ and simultaneously searching the tree of canonical paths, where the canonical path from a state to $S_0$ is found by applying the first effective token stored in the list being built.  The data stored by the algorithm consists of the current state $S$ in the search, a pointer to a token $t$ in the list of all tokens, and the list of positive tokens; we maintain as an invariant that $t$ is the first token for which we have not yet searched transition $St$.  We also maintain invariant that, whenever our search reaches a state $S$, the tokens on the canonical path from $S$ to $S_0$ have already been included in our list of positive tokens.  Initially, the list of positive tokens is empty, $S=S_0$, and $t$ is the first token in the list of tokens.

At each step of the algorithm, we test transition $St$ as follows: if $t$ is already listed in the list of positive tokens, it can not be the reverse of a step in a canonical path, and we advance $t$ in the list of all tokens.  Similarly, if $St$ is ineffective we advance $t$.  If $t$ is effective, then we search the list of tokens for the reverse token $\tilde t$ satisfying $St\tilde t=S$.  The path formed by composing the transition from $St$ back to $S$ with the canonical path from $S$ to $S_0$ must be straight (otherwise it would include $t$ and $t$ would have been already included in the list of positive tokens) so $\tilde t$ must be positive and we include it in the list of positive tokens.
We then check whether $\tilde t$ is the first listed positive token that is effective for $St$.
If so, then the transition from $S$ to $St$ is the reverse of a canonical step, so we set $S$ to $St$, reset $t$ to the beginning of the token list, output state $St$, and continue searching at the new state.  If not, we advance $t$ and continue searching from $t$.

Whenever the token $t$ advances past the end of the list of tokens in our search,
this indicates that we have exhausted all transitions from the state $S$.  So, we return to the parent of $S$ in the canonical path tree, by searching the list of positive tokens for the first one that is effective on $S$.  We must then also reset the token $t$ to the appropriate position in the list of tokens; we do this by searching sequentially through the list of tokens for the one that caused the parent of $S$ to transition to $S$.  When we advance past the end of the list of tokens for state $S_0$, the search is complete.

It is not hard to see that each step maintains the invariants discussed above.  One can prove by induction on the lengths of canonical paths that each state is output exactly once, when it is found by the reverse of a canonical step from its parent in the canonical path tree.
The space bound is easy to compute from the set of data stored by the algorithm.
For each effective transition $t$ from each state $S$, we perform $O(\tau)$ calls to the transition function:
one scan of the token list to find the reverse $\tilde t$ and determine whether $St$ is canonical, a second scan of the token list to find the effective transitions from $St$ (if $St$ was canonical), and a third scan of the token list to find the correct position of $t$ after returning from $St$ to $S$.
Therefore, the total number of calls to the transition function is $O(\tau^2)$ per state,
which dominates the total running time of the algorithm.
\end{proof}

We remark that this algorithm can easily be adapted to produce a reset sequence of length $n-1$ in the same time and space bounds: simply omit the output of each generated state, and instead output token $t$ whenever the search returns from a state $S$ to its parent $St$.

Figure~\ref{Python-reverse-search} displays an implementation of our reverse search procedure in the Python programming language~\cite{Python}.  The {\tt yield} keyword triggers Python's {\em simple generator protocol}~\cite{PEP-255}, which creates an iterator object suitable for use in {\tt for}-loops and similar contexts and returns it from each call to {\tt mediumStates}.

\section{Conclusions}

We have discussed several algorithmic problems on media, but it seems likely that some of our algorithms are non-optimal and that other interesting problems remain to be solved.

As one such problem, is it possible to efficiently find the shortest reset sequence for a medium?
Finding shortest reset sequences for general DFAs is NP-complete~\cite{Epp-SJC-90} but the reduction used to prove this does not seem to apply to media.

Another interesting direction for further research concerns random walks on media, as studied by Falmagne~\cite{Fal-JMP-97} and applied by Regenwetter et al. to electoral behavior~\cite{RegFalGro-PR-99}.  In studies of Markov systems, one of the most important parameters is the {\em mixing time}, which controls the speed at which the system reaches its stable distribution from an initial nonrandom configuration.  It is easy to construct media with mixing times that are exponential in the number of tokens: for instance, form a well-graded set family from the union of the powersets of two equal-cardinality disjoint sets; then these powersets have exponentially many members but intersect only in the empty set.  The same medium (with a random starting state) also shows that it is impossible to quickly approximate a uniform sample of a black-box medium, or to quickly list all minimal states for the content orientation of a black-box medium. However, in many natural examples of media, the mixing time seems to be small.  Is there a media-theoretic explanation for this phenomenon?

\section*{Acknowledgements}

Work of Eppstein was supported in part by NSF grant CCR-9912338.
Work of Falmagne was supported in part by NSF grant SES-9986269. 

\raggedright
\bibliographystyle{abuser}
\bibliography{media}

\end{document}